\documentstyle[aps,preprint]{revtex}

\begin{document}
\title{Path integral formulation of the tunneling dynamics of a superfluid
Fermi gas in an optical potential}
\author{M. Wouters, J. Tempere$^{\ast }$, J. T. Devreese$^{\ast \ast }$}
\address{TFVS, Universiteit Antwerpen, Universiteitsplein 1, B2610
Antwerpen, Belgium.}
\date{December 5th, 2003.}
\maketitle

\begin{abstract}
To describe the tunneling dynamics of a stack of two-dimensional fermionic
superfluids in an optical potential, we derive an effective action
functional from a path integral treatment. This effective action leads, in
the saddle point approximation, to equations of motion for the density and
the phase of the superfluid Fermi gas in each layer. In the strong coupling
limit (where bosonic molecules are formed) these equations reduce to a
discrete nonlinear Schr\"{o}dinger equation, where the molecular tunneling
amplitude is reduced for large binding energies. In the weak coupling (BCS)
regime, we study the evolution of the stacked superfluids and derive an
approximate analytical expression for the Josephson oscillation frequency in
an external harmonic potential. Both in the weak and intermediate coupling
regimes the detection of the Josephson oscillations described by our path
integral treatment constitutes experimental evidence for the fermionic
superfluid regime.
\end{abstract}

\section{Introduction}

\tighten Recent experiments have demonstrated that both degenerate Fermi
gases and Bose-Einstein condensates (BECs) can be loaded in one-dimensional
optical lattices created by standing laser waves \cite%
{AndersonScience282,CataliottiScience293,ArimondoPRL87,PhillipsJPB35,ModugnoPRA68}%
. The atoms are trapped in the valleys of the periodic potential, and form a
stack of `pancake' shaped clouds weakly coupled to each other. When the
laser power is large enough, the gases in the separate valleys become quasi
two-dimensional (quasi-2D). An additional parabolic potential, provided by
external magnetic fields, and with corresponding oscillator length much
larger than the period of the periodic potential, can be applied. In the
ground state of this system, the fermionic and/or bosonic atoms are
distributed in the lattice sites near the bottom of the additional parabolic
trap.

The possibility to load a BEC in a periodic potential has led to the
observation of the Mott-Insulator phase transition \cite{GreinerNAT415} and
the detection of Bloch oscillations and Josephson currents through the
potential barriers separating the layers of superfluid \cite%
{CataliottiScience293,ArimondoPRL87,PhillipsJPB35}. To observe the Josephson
effect, Cataliotti {\it et al.} suddenly displaced the additional parabolic
potential, placing the stack of quasi-2D BECs out of equilibrium \cite%
{CataliottiScience293}. Josephson currents allow the superfluid to tunnel
between different layers and perform pendulum-like oscillations around the
equilibrium position, driven by the external harmonic potential.

In Ref. \cite{CataliottiScience293}, a critical Josephson current was found
when the BEC was moved too far out of equilibrium, indicating the breakdown
of superfluidity across the layers \cite{MenottiNJP5}. The existence of the
Josephson currents is a direct manifestation of phase coherence across
layers. For a one-dimensional array of dilute Fermi gases the superfluid
regime is predicted to be accessible \cite{HofstetterPRL89}, and also in
this case the Josephson effect will be a signature of superfluidity.

In this paper, we derive an effective action, starting from the path
integral representation of the partition function of the coupled quasi-2D
layers filled with two different species of fermions at temperature zero.
From this effective action, we derive equations of motion that allow us to
study the dynamics of the phase and the density of the fermionic superfluid.
Based on our results for the equations of motions of a fermionic superfluid
in a one-dimensional potential, we describe what would happen if a fermionic
superfluid instead of a BEC would be subject to the experiment of Ref. \cite%
{CataliottiScience293} as illustrated in Fig. 1. We show that when the
interatomic interactions are weak there are loosely bound BCS-pairs that can
tunnel coherently through the barriers that separate the potential valleys.
When the interatomic interaction becomes stronger, confinement induced
quasi-2D bosonic molecules with high binding energy are formed and the
superfluid becomes a Bose-Einstein condensate of these molecules and we
calculate their tunneling energy.

\section{The effective action}

In the derivation of the effective action, we follow rather closely the
approach suggested by S. De Palo {\em et al. }in Ref. \cite{DePaloPRB60} and
start from the path integral representation of the partition function for a
system consisting of layers of 2D fermions%
\begin{equation}
Z=\int {\cal D}\psi _{j,\sigma }^{\dag }(x){\cal D}\psi _{j,\sigma }(x)\exp
\left\{ -S\left[ \psi _{j,\sigma }^{\dag }(x),\psi _{j,\sigma }(x)\right]
\right\} ,
\end{equation}%
where the action is given by%
\begin{eqnarray}
S\left[ \psi _{j,\sigma }^{\dag }(x),\psi _{j,\sigma }(x)\right] &=&\sum_{j}%
\displaystyle\int %
\limits_{0}^{\beta }d\tau 
\displaystyle\int 
d^{2}{\bf x}\left[ \sum_{\sigma }\psi _{j,\sigma }^{\dag }\left( x\right)
\left( \partial _{\tau }-\frac{\nabla ^{2}}{2m}+V_{ext}\left( j\right) -\mu
\right) \psi _{j,\sigma }\left( x\right) \right.  \nonumber \\
&&-U\psi _{j,\uparrow }^{\dag }\left( x\right) \psi _{j,\downarrow }^{\dag
}\left( x\right) \psi _{j,\downarrow }\left( x\right) \psi _{j,\uparrow
}\left( x\right)  \nonumber \\
&&\left. +t_{1}\sum_{\sigma }\left( \psi _{j,\sigma }^{\dag }\left( x\right)
\psi _{j+1,\sigma }\left( x\right) +\psi _{j+1,\sigma }^{\dag }\left(
x\right) \psi _{j,\sigma }\left( x\right) \right) \right] .  \label{S}
\end{eqnarray}%
Here, $\beta =1/(k_{B}T)$ where $T$ denotes the temperature and $k_{B}$ the
Boltzmann constant. The three-vector notation $x=\left( {\bf x,}\tau \right) 
$ is used. The field $\psi _{j,\sigma }\left( x\right) $ belongs to a
fermion of mass $m$ in layer $j$ and spin $\sigma $ ($\uparrow $ or $%
\downarrow $). The potential $V_{ext}\left( j\right) $ is an additional
external potential that the fermions are subjected to. The attraction
strength between the fermions is determined by $U$. The interlayer tunneling
energy for a fermion is denoted by the real number $t_{1}$ that can be
calculated with the approximate formula from Ref. \cite%
{martikainencondmat0304} for the tunneling energy in a optical potential $%
V\left( z\right) =V_{0}\sin ^{2}\left( 2\pi z/\lambda \right) $ with wave
length $\lambda $ and depth $V_{0}$:%
\begin{equation}
t_{1}=\frac{m\omega _{L}^{2}\lambda ^{2}}{8\pi ^{2}}\left[ \frac{\pi ^{2}}{4}%
-1\right] e^{-\left( \lambda /4\ell _{L}\right) ^{2}},  \label{t1}
\end{equation}%
where $\omega _{L}=\sqrt{8\pi ^{2}V_{0}/\left( m\lambda ^{2}\right) }$ and $%
\ell _{L}=\sqrt{1/m\omega _{L}}$ are respectively the trapping frequency and
the oscillator length that an atom feels in in de $z$-direction.

In appendix A, the reduction of (\ref{S}) to an effective action is given
and here we only give a summary. In order to grasp the most important part
of the path integral in the superfluid state, the interaction between the
fermions is decoupled by the Hubbard-Stratonovich (HS) transformation with
the complex HS-field $\Delta _{j}^{HS}(x)$. After integration over the
fermion fields, one is left with an effective action in terms of the
HS-fields.

However, no information about the physical density of the system can be read
off from such an effective action. In order have access to this quantity in
the effective action, we introduce it by multiplying the partition function
with the constant%
\begin{eqnarray}
C &=&\int {\cal D}\zeta _{j}^{HS}(x){\cal D}\rho _{j}(x)\exp \left\{
-\sum_{j}%
\displaystyle\int %
\limits_{0}^{\beta }d\tau \int d^{2}{\bf x}\;i\zeta _{j}^{HS}\left( x\right)
\right.  \nonumber \\
&&\left. \times \left[ \rho _{j}\left( x\right) -\psi _{j,\uparrow }^{\dag
}\left( x\right) \psi _{j,\uparrow }\left( x\right) -\psi _{j,\downarrow
}^{\dag }\left( x\right) \psi _{j,\downarrow }\left( x\right) \right]
\right\} .  \label{eqC}
\end{eqnarray}%
Carrying out the functional integral over $\zeta _{j}^{HS}\left( x\right) $
alone gives $\delta \lbrack \rho _{j}\left( x\right) -\psi _{j,\uparrow
}^{\dag }\left( x\right) \psi _{j,\uparrow }\left( x\right) -\psi
_{j,\downarrow }^{\dag }\left( x\right) \psi _{j,\downarrow }\left( x\right)
]$ and thus $\rho _{j}\left( x\right) $ corresponds to the physical density
of the system along any path. Next, the complex field $\Delta _{j}^{HS}(x)$
is separated in a modulus and a phase. This phase is important for the low
energy dynamics and therefore it is advantageous to introduce it explicitly 
\begin{equation}
\Delta _{j}^{HS}\left( x\right) =\left| \Delta _{j}^{HS}\left( x\right)
\right| e^{i\theta _{j}\left( x\right) }.  \label{deltaexplicit}
\end{equation}%
We then arrive at the following expression for the partition function 
\begin{equation}
Z\propto \int {\cal D}\left| \Delta _{j}^{HS}\left( x\right) \right| {\cal D}%
\zeta _{j}^{HS}(x){\cal D}\theta _{j}(x){\cal D}\rho _{j}(x)\exp \left[ -S_{%
\text{eff}}\right] ,
\end{equation}%
where $S_{\text{eff}}$ is given in appendix A, expression (\ref{Seff}).

\bigskip

To describe the low-energy dynamics of the density and the phase of the
superfluid, the paths along which $\theta _{j}(x)$ and $\rho _{j}(x)$ vary
slowly in comparison to the fermionic frequencies (Fermi energy and binding
energy) will be of importance. Along these paths, we make a saddle point
approximation for the remaining fields $\left| \Delta _{j}^{HS}(x)\right| $
and $\zeta _{j}^{HS}(x)$ in appendix B. The fluctuations $\delta \left|
\Delta _{j}^{HS}(x)\right| $ and $\delta \zeta _{j}^{HS}(x)$ around the
saddle point values $\left| \Delta _{j}^{(0)}(x)\right| $ and $\zeta
_{j}^{(0)}(x)$ can be treated perturbatively. The saddle point value for the
effective action is calculated in the appendix, and given by expression (\ref%
{Seffsp}). The saddle point equations are 
\begin{eqnarray}
\frac{1}{U} &=&\int \frac{d^{2}{\bf k}}{\left( 2\pi \right) ^{2}}\frac{%
1-2n_{F}\left[ E_{j}(k)\right] }{2E_{j}\left( k\right) },  \label{egap0} \\
\rho _{j}(x) &=&\int \frac{d^{2}{\bf k}}{\left( 2\pi \right) ^{2}}\left( 
\frac{k^{2}}{2m}-i\zeta _{j}^{(0)}(x)\right) \left\{ \frac{2n_{F}\left[
E_{j}(k)\right] -1}{E_{j}\left( k\right) }+1\right\} ,  \label{edens}
\end{eqnarray}%
with $n_{F}(E)=1/(e^{\beta E}+1)$ the Fermi-Dirac distribution function and $%
E_{j}\left( k\right) $ the local BCS energy defined by%
\begin{equation}
E_{j}\left( k\right) =\sqrt{\left( \frac{k^{2}}{2m}-i\zeta
_{j}^{(0)}(x)\right) +\left| \Delta _{j}^{(0)}(x)\right| ^{2}}.
\label{BCSenergy}
\end{equation}%
Equations (\ref{edens}) and (\ref{BCSenergy}) show that the saddle point
value $\zeta _{j}^{(0)}(x)$ can be interpreted as a chemical potential $%
z_{j}(x)=i\zeta _{j}^{(0)}(x)$. The first saddle point equation (\ref{egap0}%
) corresponds to the BCS gap equation, whereas the second saddle point
equation leads to the BCS equation fixing the chemical potential $z_{j}(x)$
in layer $j$ as a function of the density $\rho _{j}(x)$ in layer $j$.

As we have introduced a momentum independent contact-interaction, equation (%
\ref{egap0}) has to be regularized as described in \cite{houbiersPRA56},
after which it becomes%
\begin{equation}
\frac{-1}{T_{00}\left( E\right) }=\int \frac{d^{2}{\bf k}}{\left( 2\pi
\right) ^{2}}\left[ \frac{2n_{F}\left[ E_{j}(k)\right] -1}{2E_{j}\left(
k\right) }-\frac{1}{k^{2}/m-E+i\varepsilon }\right] ,  \label{egap}
\end{equation}%
where $T_{00}\left( E\right) $ is the low-momentum limit of the $T$-matrix.
This equation has no ultraviolet divergences. In two dimensions, at low
energy, $T_{00}\left( E\right) $ is given by \cite%
{RanderiaPRB41,AdhikariAmJPhys54} 
\begin{equation}
\frac{1}{T_{00}\left( E\right) }=\frac{m}{4}\left[ \frac{-1}{\pi }\ln \left[
E/E_{b}\right] +i\right] ,  \label{Tmatrix}
\end{equation}%
where $E_{b}$ is the energy of the 2D bound state that always exists in two
dimensions. For the optical potential $V_{0}\sin ^{2}\left( 2\pi z/\lambda
\right) $ described earlier, the binding energy of the quasi-2D bound state
is given by \cite{petrovPRA64}%
\begin{equation}
E_{b}=\frac{C\hbar \omega _{L}}{\pi }\exp \left( \sqrt{2\pi }\frac{\ell _{L}%
}{a}\right) ,  \label{Ebind}
\end{equation}%
with $a$ the scattering length of the fermionic atoms and $C\approx 0.915.$

\section{Josephson current at $T=0$}

\subsection{Equations of motion for density and phase}

We now proceed with an analysis at $T=0$ and with the assumption that the
energy $t_{1}$ is small compared to the other energies, such that a
perturbational expansion with $t_{1}$ as a small parameter is possible. In
this case, equations (\ref{edens}) and (\ref{egap}) can be solved
analytically for $\left\vert \Delta _{j}^{(0)}(x)\right\vert $ and $z_{j}(x)$
(see also \cite{RanderiaPRB41}) :%
\begin{eqnarray}
\left\vert \Delta _{j}^{(0)}(x)\right\vert &=&\sqrt{\frac{2\pi \rho _{j}(x)}{%
m}E_{b}},  \label{sp1} \\
z_{j}(x) &=&\frac{\pi \rho _{j}(x)}{m}-\frac{E_{b}}{2}.  \label{sp2}
\end{eqnarray}

As we want to study the current perpendicular to the layers in which the
atoms are confined, we have to calculate the terms in the effective action
that couple the different layers. In our perturbational expansion of the
effective action (\ref{Seffsp}), the lowest order contribution comes from
the term in the self energy (\ref{Sigma}) proportional to $t_{1}$. We find
that the contribution equals%
\begin{eqnarray}
&&-2t_{1}^{2}\sum_{j}\frac{1}{\beta }\sum_{\omega _{n}=(2n+1)\pi /\beta
}\int \frac{d^{2}{\bf k}}{\left( 2\pi \right) ^{2}}\frac{1}{\left[ \omega
_{n}^{2}+E_{j+1}^{2}\left( k\right) \right] \left[ \omega
_{n}^{2}+E_{j}^{2}\left( k\right) \right] } \\
&&\times \left\{ -\omega _{n}^{2}+\xi _{j+1}\left( k\right) \xi _{j}\left(
k\right) +\left\vert \Delta _{j+1}^{(0)}(x)\right\vert \left\vert \Delta
_{j}^{(0)}(x)\right\vert \cos \left[ \theta _{j+1}(x)-\theta _{j}(x)\right]
\right\} ,
\end{eqnarray}%
with $\xi _{j}\left( k\right) =k^{2}/\left( 2m\right) -z_{j}(x)$ the free
fermion dispersion and $E_{j}$ the BCS dispersion relation (\ref{BCSenergy}%
). In this expression, the part proportional to $\cos \left[ \theta
_{j+1}(x)-\theta _{j}(x)\right] $ is responsible for the Josephson tunneling
between the layers and we write it symbolically as 
\begin{equation}
S_{\text{tunnel}}=-\sum_{j}\text{ }\int_{0}^{\beta }d\tau \int d{\bf x}\text{
}T_{j+1,j}\cos \left[ \theta _{j+1}(x)-\theta _{j}(x)\right] .
\label{Stunnel}
\end{equation}%
Evaluating this term with the supposition that the gap $\left\vert \Delta
_{j}^{(0)}(x)\right\vert $ and chemical potential $z_{j}(x)$ vary slowly
with $j$ leads to%
\begin{equation}
T_{j+1,j}=\frac{t_{1}^{2}m}{4\pi }\left( 1+\frac{z_{j}(x)}{\sqrt{\left\vert
\Delta _{j}^{(0)}(x)\right\vert ^{2}+z_{j}^{2}(x)}}\right) =\frac{%
t_{1}^{2}\rho _{j}(x)}{2\pi \rho _{j}(x)/m+E_{b}}.
\end{equation}%
A similar contribution to the energy can be obtained in a BCS-approach,
following Ref. \cite{TanakaphysicaC219}.

We now take as an approximation for the effective action 
\begin{equation}
S_{\text{J}}\left[ \theta _{j}(x),\rho _{j}(x)\right] =S_{\text{eff}}^{\text{%
sp}}+S_{\text{tunnel}},
\end{equation}%
with $S_{\text{eff}}^{\text{sp}}$ given by (\ref{Seffsp})\ from which $%
\left| \Delta _{j}^{0}(x)\right| $ and $\zeta _{j}^{0}(x)$ have been
eliminated using equations (\ref{sp1}),(\ref{sp2}). Having obtained an
expression for the action that only depends on $\theta _{j}(x)$ and $\rho
_{j}(x)$ we can finally proceed with deriving equations of motion by
extremizing $S_{\text{J}}$ with respect to these fields. Because we want to
give a dynamical interpretation to these equations, we write them in real
time ($i\partial _{\tau }\rightarrow \partial _{t}$). Extremizing with
respect to the phase field $\theta _{j}(x)$ results in 
\begin{eqnarray}
\partial _{t}\frac{\rho _{j}\left( x\right) }{2} &=&-\frac{{\bf \nabla }%
\theta _{j}\left( x\right) \cdot {\bf \nabla }\rho _{j}\left( x\right) }{4m}
\nonumber \\
&&+T_{j,j-1}\sin \left[ \theta _{j}(x)-\theta _{j-1}(x)\right]  \nonumber \\
&&-T_{j+1,j}\sin \left[ \theta _{j+1}(x)-\theta _{j}(x)\right] ,
\label{eqcont}
\end{eqnarray}%
and the derivative with respect to $\rho _{j}\left( x\right) $ yields%
\begin{eqnarray}
-\partial _{t}\frac{\theta _{j}\left( x\right) }{2} &=&\frac{\left[ \nabla
\theta _{j}\left( x\right) \right] ^{2}}{8m}+V_{ext}\left( j\right)
+z_{j}-\mu  \nonumber \\
&&-\frac{\partial T_{j+1,j}}{\partial \rho _{j}\left( x\right) }\cos \left[
\theta _{j+1}(x)-\theta _{j}(x)\right]  \nonumber \\
&&-\frac{\partial T_{j,j-1}}{\partial \rho _{j}\left( x\right) }\cos \left[
\theta _{j}(x)-\theta _{j-1}(x)\right] .  \label{eEuler}
\end{eqnarray}%
We have calculated the derivative $\partial T_{j+1,j}/\partial \rho _{j}$ if
the density varies smoothly with the layer index ($\rho _{j+1}\approx \rho
_{j}\approx \rho _{j-1}$) and is constant in the plane:%
\begin{equation}
\frac{\partial T_{j+1,j}}{\partial \rho _{j}(x)}=\frac{t_{1}^{2}}{2}\frac{%
E_{b}}{\left[ 2\pi \rho _{j}(x)/m+E_{b}\right] ^{2}}.
\end{equation}

\bigskip

\subsection{Oscillations of the superfluid in an optical lattice}

We now introduce the wave function 
\begin{equation}
\psi _{j}(t)=\sqrt{\frac{\rho _{j}(t)}{2}}e^{i\theta _{j}(t)},
\label{wavfie}
\end{equation}%
with $\rho _{j}$ and $\theta _{j}$ only depending on time and layer index $j$%
. That is, we assume that within a layer, the density and the phase are
homogeneous, but that they can still vary over time and over layers. The
wave function (\ref{wavfie}) obeys the Schr\"{o}dinger equation%
\begin{eqnarray}
i\frac{d}{dt}\psi _{j} &=&\left( 2V_{ext}\left( j\right) +2z_{j}-2\mu
\right) \psi _{j}-\frac{t_{1}^{2}}{2\pi \rho _{j}/m+E_{b}}\psi _{j}\left(
e^{i\left( \theta _{j-1}-\theta _{j}\right) }+e^{i\left( \theta
_{j+1}-\theta _{j}\right) }\right)  \nonumber \\
&&-\psi _{j}t_{1}^{2}\frac{4\pi \rho _{j}/m}{\left( E_{b}+2\pi \rho
_{j}/m\right) ^{2}}\left[ \cos \left( \theta _{j+1}-\theta _{j}\right) +\cos
\left( \theta _{j}-\theta _{j-1}\right) \right] ,  \label{eqpsitot}
\end{eqnarray}%
where we can take in the approximation of slowly varying density that we
used before, namely $\psi _{j-1}\approx \psi _{j}e^{i\left( \theta
_{j-1}-\theta _{j}\right) }$ and $\psi _{j+1}\approx \psi _{j}e^{i\left(
\theta _{j+1}-\theta _{j}\right) }$, so that we have in the limit $\pi \rho
_{j}/m\ll E_{b}$ that 
\begin{equation}
i\frac{d}{dt}\psi _{j}=\left( 2V_{ext}\left( j\right) +2z_{j}-2\mu \right)
\psi _{j}-\frac{t_{1}^{2}}{2\pi \rho _{j}/m+E_{b}}\left( \psi _{j-1}+\psi
_{j+1}\right) ,  \label{eqpsi}
\end{equation}%
which is the nonlinear discrete Schr\"{o}dinger equation for an array of
Bose-Einstein condensates \cite{trombettoniPRL86}, where the boson tunneling
matrix element is given by $t_{1}^{2}/\left( 2\pi \rho _{j}/m+E_{b}\right) $%
. This is a decreasing function of $E_{b}$, which has the physical
consequence that for increasing molecular binding energy the coupling
between the layers becomes weaker and that in the Bose-Einstein limit ($%
E_{b}\rightarrow \infty $), no tunneling of molecules occurs. This may be
related to the fact that the binding together of atoms in molecules is
strongly influenced by the the confinement potential and the quasi-2D nature
of the gas (as can be seen from eq. \ref{Ebind}). In the intermediate states
in the calculation of the amplitude of the tunneling process, the molecular
state is unlikely to survive as a bound state. Hence, the molecular binding
energy is to be added to the energy barrier for tunneling.

We know from \cite{CataliottiScience293}, that {\it in the bosonic limit}
equation (\ref{eqpsi}) allows for oscillations where the phase difference $%
\theta _{j+1}-\theta _{j}$ is locked to a constant value $\Delta \theta $
and the differential equation governing the dynamics of the center of mass
coordinate and this phase difference $\Delta \theta $ is of the pendulum
type. Our numerical simulations of (\ref{eqpsitot}) suggest that {\it in the
BCS-regime} there is a similar current through the array.

>From equations (\ref{eqcont}) and (\ref{eEuler}), we can derive a
simplified analytical formula to estimate the oscillation frequency in an
external harmonic potential $V_{ext}\left( j\right) =\Omega j^{2}$. Assuming
that the phase difference between neighboring layers is constant, this
becomes%
\begin{equation}
\partial _{t}R=2\left\langle T_{j,j-1}\right\rangle \sin \left( \theta
_{j}-\theta _{j-1}\right) ,  \label{esimp1}
\end{equation}%
where 
\[
\left\langle T_{j,j-1}\right\rangle =\frac{1}{\sum_{j}\rho _{j}\left(
x\right) }\sum_{j}T_{j,j-1} 
\]%
is the average of $T_{j,j-1}$ over the lattice sites. The initial density
profile $\rho _{j}(t=0)$ is calculated from the chemical potentials $z_{j}$
through (\ref{sp2}) and these chemical potentials are derived from (\ref%
{eEuler}), $z_{j}=\mu -V_{ext}\left( j\right) $, for a particular choice of $%
\mu $. The last two terms from equation (\ref{eqpsitot}) seem to have only a
minor influence on the frequency so that we omit them in the analytic
calculation. Equation (\ref{eEuler}) then becomes%
\begin{equation}
\partial _{t}\left( \theta _{j}-\theta _{j-1}\right) =-4\Omega R.
\label{esimp2}
\end{equation}%
For small oscillations, (\ref{esimp1}) and (\ref{esimp2}) lead to the
frequency 
\begin{equation}
\omega =\sqrt{8\Omega \left\langle T_{j,j-1}\right\rangle }.  \label{wanal}
\end{equation}

Taking $^{40}$K atoms with a central density of $n=10^{9}$ cm$^{-2},$ an
optical wavelength $\lambda =754$ nm, and an axial frequency $\omega
_{a}=2\pi \times 24$ Hz, we plot in Fig. 2 the analytical frequency (curves)
obtained from equation (\ref{wanal}) and compare it with a numerical
calculation (symbols) based on equation (\ref{eqpsitot}). The oscillation
frequency is plotted as a function of the inverse scattering length which
appears in expression (\ref{Ebind}) of the binding energy. Fig. 2 shows that
the estimation (\ref{wanal}) agrees almost perfectly with the numerical
results in the bosonic limit, and that also in the BCS regime there is
reasonable quantitative agreement. The inset of Fig. 2 shows that far in the
bosonic regime ($a>0$), the oscillation frequency decreases rapidly as an
exponential function of $1/a$. Nevertheless in the cross-over regime, the
oscillation frequencies are high enough for the observation of Josephson
currents to be useful as a tool to investigate the superfluidity of an
ultracold Fermi gas, analogous to the experiments of Ref. \cite%
{CataliottiScience293} in the bosonic case.

\section{Conclusions}

We have derived an effective action and the resulting equations of motion to
describe the dynamics of a fermionic superfluid in a layered system and
applied this formalism to study the center of mass motion of an atomic Fermi
gas in the potential formed by an optical standing wave. We find that a
Fermi gas {\it in the BCS regime }can perform superfluid oscillations
through the optical lattice, similar to those that have been observed for
condensates of bosonic atoms \cite%
{CataliottiScience293,ArimondoPRL87,PhillipsJPB35}, when the gas is not in
equilibrium in the harmonic trapping potential superimposed on the optical
lattice. An analytical approximate expression (\ref{wanal}) for the
oscillation frequency is derived and the predictions of this expression are
tested with numerical simulations of the full equations of motion (\ref%
{eqpsitot}). For a superfluid Fermi gas {\it in the BEC regime}, we find
that the tunneling is suppressed when the molecular binding energy becomes
large. We conclude that superfluidity in Fermi gases can be revealed through
Josephson currents in optical lattices if the Fermi gas is either in the BCS
regime, or in the weakly-bound molecular BEC regime.

\section{Acknowledgments}

We thank G. Modugno for useful discussions. Two of the authors (M. W. and J.
T.) are supported financially by the Fund for Scientific Research - Flanders
(Fonds voor Wetenschappelijk Onderzoek -- Vlaanderen). This research has
been supported financially by the FWO-V projects Nos. G.0435.03, G.0306.00,
the W.O.G. project WO.025.99N.and the GOA BOF UA 2000, IUAP.

\appendix%

\section{Derivation of the effective action}

In order to grasp the most important part of the path integral in the
paired-fermion state, the interaction between the fermions is decoupled by
the Hubbard-Stratonovich transformation, which transforms (\ref{S}) into 
\begin{equation}
Z=\int {\cal D}\psi _{j,\sigma }^{\dag }(x){\cal D}\psi _{j,\sigma }(x){\cal %
D}\Delta _{j}^{HS}(x){\cal D}\Delta _{j}^{HS,\dagger }(x)\exp \left\{
-S^{\left( 1\right) }\right\} ,
\end{equation}%
with%
\begin{eqnarray}
S^{\left( 1\right) } &=&\sum_{j}\int_{0}^{\beta }d\tau \int d^{2}{\bf x}%
\left[ \frac{\left\vert \Delta _{j}^{HS}\left( x\right) \right\vert ^{2}}{U}%
\right.  \nonumber \\
&&+\sum_{\sigma =\pm 1}\psi _{j,\sigma }^{\dag }\left( x\right) \left(
\partial _{\tau }-\frac{\nabla ^{2}}{2m}+V_{ext}\left( j\right) -\mu \right)
\psi _{j,\sigma }\left( x\right)  \nonumber \\
&&-\Delta _{j}^{HS}\left( x\right) \psi _{j,\uparrow }^{\dag }\left(
x\right) \psi _{j,\downarrow }^{\dag }\left( x\right) -\Delta _{j}^{HS,\dag
}\left( x\right) \psi _{j,\downarrow }\left( x\right) \psi _{j,\uparrow
}\left( x\right)  \nonumber \\
&&+\left. t_{1}\sum_{\sigma }\left( \psi _{j,\sigma }^{\dag }\left( x\right)
\psi _{j+1,\sigma }\left( x\right) +\psi _{j+1,\sigma }^{\dag }\left(
x\right) \psi _{j,\sigma }\left( x\right) \right) \right] .
\end{eqnarray}%
In order to keep track of the total density, we introduce the constant $C$
(see expression (\ref{eqC})). In order to investigate the BCS gap and the
phase we separate the complex field $\Delta _{j}^{HS}(x)$ in a modulus and a
phase (expression (\ref{deltaexplicit}))\ and also transform the fermion
fields as $\psi _{j,\sigma }\left( x\right) \rightarrow \psi _{j,\sigma
}\left( x\right) e^{i\theta _{j}\left( x\right) /2}$. Additionally, we shift
the field $i\zeta _{j}^{HS}\left( x\right) $ according to 
\begin{equation}
i\zeta _{j}^{HS}\left( x\right) \rightarrow i\zeta _{j}^{HS}\left( x\right)
+i\partial _{\tau }\frac{\theta _{j}\left( x\right) }{2}+\frac{\left( \nabla
\theta _{j}\left( x\right) \right) ^{2}}{8m}+V_{ext}\left( j\right) -\mu
\end{equation}%
and use the Nambu spinor notation $\eta _{j}\left( x\right) =\left( \psi
_{j,\uparrow }\left( x\right) ,\psi _{j,\downarrow }^{\dag }\left( x\right)
\right) ^{T}$. After this procedure, the partition function can be written
as 
\begin{equation}
Z\propto \int \left\vert \Delta _{j}^{HS}(x)\right\vert {\cal D}\eta
_{j}^{+}(x){\cal D}\eta _{j}(x){\cal D}\left\vert \Delta
_{j}^{HS}(x)\right\vert {\cal D}\theta _{j}(x){\cal D}\zeta _{j}^{HS}(x)%
{\cal D}\rho _{j}(x)\exp \left\{ -S^{\left( 2\right) }\right\} ,
\end{equation}%
with the action $S^{\left( 2\right) }=S_{0}+S^{(3)}[\eta _{j}^{+}(x),\eta
_{j}(x)]$ where%
\begin{eqnarray}
S_{0} &=&\sum_{j}\int_{0}^{\beta }d\tau \int d^{2}{\bf x}\left\{ \frac{%
\left\vert \Delta _{j}^{HS}(x)\right\vert ^{2}}{U}\right.  \nonumber \\
&&\left. +\left[ i\zeta _{j}^{HS}(x)+i\partial _{\tau }\frac{\theta
_{j}\left( x\right) }{2}+\frac{\left( \nabla \theta _{j}\left( x\right)
\right) ^{2}}{8m}+V_{ext}\left( j\right) -\mu \right] \rho _{j}\left(
x\right) \right\}  \label{Snul}
\end{eqnarray}%
does not contain the fermion fields any more and 
\begin{eqnarray}
S^{(3)} &=&\sum_{j}\int_{0}^{\beta }d\tau \int d^{2}{\bf x}\left\{ \eta
_{j}^{\dag }\left( x\right) \left[ \left( \partial _{\tau }-i\frac{\nabla
\theta _{j}\left( x\right) }{2m}\nabla -i\frac{\nabla ^{2}\theta _{j}\left(
x\right) }{4m}\right) \sigma _{0}\right. \right.  \nonumber \\
&&+\left. \left( -\frac{\nabla ^{2}}{2m}-i\zeta _{j}^{HS}\left( x\right)
\right) \sigma _{3}-\left\vert \Delta _{j}^{HS}\left( x\right) \right\vert
\sigma _{1}\right] \eta _{j}\left( x\right)  \nonumber \\
&&\left. +\left[ t_{1}\eta _{j}^{\dag }\left( x\right) e^{i\left( \theta
_{j+1}-\theta _{j}\right) \sigma _{3}/2}\sigma _{3}\eta _{j+1}\left(
x\right) +\text{h.c.}\right] \right\}
\end{eqnarray}%
is the part of the action that still depends on them. The Pauli matrices are
denoted by $\sigma _{i}$. Since $S^{(3)}$ is quadratic in the fermion
fields, the path integral over these fields can be performed, resulting in 
\begin{equation}
\int {\cal D}\eta _{j}^{+}(x){\cal D}\eta _{j}(x)\exp \{-S^{(3)}\}=\det
[-G^{-1}],
\end{equation}%
where the Green's function $G$ is a matrix in coordinate space as in layer
space and is given by%
\begin{eqnarray}
-G^{-1}\left( x,j;x^{\prime },j^{\prime }\right) &=&\delta \left(
x-x^{\prime }\right) \left\{ \delta _{jj^{\prime }}\left[ \left( \partial
_{\tau }-i\frac{\nabla \theta _{j}\left( x\right) }{2m}\nabla -i\frac{\nabla
^{2}\theta _{j}\left( x\right) }{4m}\right) \sigma _{0}\right. \right. 
\nonumber \\
&&\left. +\left( -\frac{\nabla ^{2}}{2m}-i\zeta _{j}^{HS}(x)\right) \sigma
_{3}-\left\vert \Delta _{j}^{HS}(x)\right\vert \sigma _{1}\right]  \nonumber
\\
&&\left. +\delta _{j+1,j^{\prime }}t_{1}e^{i\left( \theta _{j+1}-\theta
_{j}\right) \sigma _{3}/2}\sigma _{3}+\delta _{j-1,j^{\prime
}}t_{1}e^{-i\left( \theta _{j+1}-\theta _{j}\right) \sigma _{3}/2}\sigma
_{3}\right\} .  \label{Greens}
\end{eqnarray}%
The partition sum can then be written as 
\begin{equation}
Z\propto \int {\cal D}\left\vert \Delta _{j}^{HS}(x)\right\vert {\cal D}%
\theta _{j}(x){\cal D}\zeta _{j}^{HS}(x){\cal D}\rho _{j}(x)\exp \left\{ -S_{%
\text{eff}}\right\}  \label{Z4}
\end{equation}%
where 
\begin{equation}
S_{\text{eff}}=S_{0}+\text{Tr}\left[ \ln (-G^{-1})\right]  \label{Seff}
\end{equation}%
with $S_{0}$ given by (\ref{Snul}) and the Green's function given by (\ref%
{Greens}).

\section{Saddle point expansion for the effective action}

>From the path integrations in (\ref{Z4}), we ultimately want to extract
equations of motion for $\theta _{j}(x),$ $\rho _{j}(x)$. To describe the
low-energy dynamics, the contributions with $\theta _{j}(x),$ $\rho _{j}(x)$
varying slow in comparison with the fermionic frequencies (Fermi energy and
binding energy) will be important. Along the paths of slowly varying $\theta
_{j}(x),$ $\rho _{j}(x)$, we can make the saddle point expansion in the
fields $\left| \Delta _{j}^{HS}\right| ,\zeta _{j}^{HS}$, setting%
\begin{eqnarray}
\left| \Delta _{j}^{HS}(x)\right| &=&\left| \Delta _{j}^{(0)}\left( x\right)
\right| +\delta \left| \Delta _{j}^{HS}(x)\right| , \\
\zeta _{j}^{HS}(x) &=&\zeta _{j}^{(0)}\left( x\right) +\delta \zeta
_{j}^{HS}(x),
\end{eqnarray}%
where also $\left| \Delta _{j}^{(0)}\left( x\right) \right| $ and $\zeta
_{j}^{(0)}\left( x\right) $ will vary slowly in comparison to the fermion
frequencies.

Expanding the Green's function (\ref{Greens}) around the saddle point values 
$\left| \Delta _{j}^{(0)}\right| $ and $\zeta _{j}^{(0)}=-iz_{j}$ leads to 
\begin{equation}
G^{-1}=G_{0}^{-1}+\Sigma ,
\end{equation}%
where the saddle point contribution is given by 
\begin{equation}
-G_{0}^{-1}\left( x,j;x^{\prime },j^{\prime }\right) =\delta \left(
x-x^{\prime }\right) \delta _{jj^{\prime }}\left[ \partial _{\tau }\sigma
_{0}+\left( -\frac{\nabla ^{2}}{2m}-z_{j}\right) \sigma _{3}-\left| \Delta
_{j}^{(0)}\right| \sigma _{1}\right] .
\end{equation}%
This saddle point contribution can be diagonalized by going to Fourier space
: 
\begin{equation}
G_{0}^{-1}\left( {\bf k,}\omega ,j;{\bf k}^{\prime },\omega ^{\prime
},j^{\prime }\right) =\delta _{jj^{\prime }}\delta _{\omega \omega ^{\prime
}}\delta \left( {\bf k}-{\bf k}^{\prime }\right) \left[ i\omega \sigma
_{0}-\left( \frac{k^{2}}{2m}-z_{j}\right) \sigma _{3}+\left| \Delta
_{j}^{(0)}\right| \sigma _{1}\right] .
\end{equation}%
and thus%
\begin{eqnarray}
\mathop{\rm Tr}%
\left[ \ln \left[ -G_{0}^{-1}\right] \right] &=&\sum_{j}\frac{1}{\beta }%
\sum_{\omega _{n}=(2n+1)\pi /\beta }\int \frac{d^{2}{\bf k}}{(2\pi )^{2}}%
\text{ }  \nonumber \\
&&\times \ln \left[ -\omega _{n}^{2}-\left[ k^{2}/(2m)-z_{j}(x)\right]
^{2}+\left| \Delta _{j}^{(0)}(x)\right| ^{2}\right] .  \label{G0}
\end{eqnarray}%
In this expression, the $x$ dependence of $z_{j}(x)$ and $\left| \Delta
_{j}^{(0)}(x)\right| $ has been reintroduced, still assuming that these
fields vary slowly in comparison to the relevant fermion frequencies. The
self energy $\Sigma $ equals%
\begin{eqnarray}
-\Sigma \left( x,j;x^{\prime },j^{\prime }\right) &=&\delta \left(
x-x^{\prime }\right) \delta _{jj^{\prime }}\left[ -i\sigma _{3}\delta \zeta
_{j}^{HS}(x)-(\delta \left| \Delta _{j}^{HS}(x)\right| )\sigma _{1}\right. 
\nonumber \\
&&\left. -\left( i\frac{\nabla \theta _{j}}{2m}\nabla +i\frac{\nabla
^{2}\theta _{j}}{4m}\right) \sigma _{0}\right]  \nonumber \\
&&+\delta \left( x-x^{\prime }\right) \left[ \delta _{j+1,j^{\prime
}}t_{1}e^{i\left[ \theta _{j+1}(x)-\theta _{j}(x)\right] \sigma
_{3}/2}\sigma _{3}\right.  \nonumber \\
&&\left. +\delta _{j-1,j^{\prime }}t_{1}e^{-i\left[ \theta _{j+1}(x)-\theta
_{j}(x)\right] \sigma _{3}/2}\sigma _{3}\right] ,  \label{Sigma}
\end{eqnarray}%
so that%
\begin{eqnarray}
\det \left[ -G^{-1}\right] &=&\exp \left\{ 
\mathop{\rm Tr}%
\left[ \ln \left( -G^{-1}\right) \right] \right\} =\exp \left\{ 
\mathop{\rm Tr}%
\left[ \ln \left( -G_{0}^{-1}-\Sigma \right) \right] \right\}  \nonumber \\
&=&\exp \left\{ 
\mathop{\rm Tr}%
\left[ \ln \left( -G_{0}^{-1}\right) \right] -\sum_{n=1}^{\infty }\frac{%
\mathop{\rm Tr}%
\left[ \left( -G_{0}^{-1}\Sigma \right) ^{n}\right] }{n}\right\} ,  \label{G}
\end{eqnarray}%
where $%
\mathop{\rm Tr}%
$ denotes the trace over all the variables (coordinates, imaginary time,
layer index and Nambu space). Expanding the action, $S_{\text{eff}}$,
expression (\ref{Seff}), around the saddle point values $\left| \Delta
_{j}^{(0)}(x)\right| $ and $z_{j}(x)=i\zeta _{j}^{(0)}(x)$, and using the
result (\ref{G}) to lowest order, we find 
\begin{eqnarray}
S_{\text{eff}}^{\text{sp}} &=&%
\mathop{\rm Tr}%
\left[ \ln \left( -G_{0}^{-1}\right) \right] +\sum_{j}\int_{0}^{\beta }d\tau
\int d^{2}{\bf x}\left\{ \frac{\left| \Delta _{j}^{(0)}(x)\right| ^{2}}{U}%
\right.  \nonumber \\
&&\left. +\left[ z_{j}(x)+i\partial _{\tau }\frac{\theta _{j}\left( x\right) 
}{2}+\frac{\left( \nabla \theta _{j}\left( x\right) \right) ^{2}}{8m}%
+V_{ext}\left( j\right) -\mu \right] \rho _{j}\left( x\right) \right\}
\label{Seffsp}
\end{eqnarray}%
The saddle point values $\left| \Delta _{j}^{\left( 0\right) }(x)\right| $
and $z_{j}(x)=i\zeta _{j}^{(0)}(x)$ are now equal to the extremum of (\ref%
{Seffsp}) and they fulfill the equations (\ref{egap0}) and (\ref{edens}).
The lowest order Green's function $G_{0}$ is given by (\ref{G0}).

\bigskip

\bigskip 
\begin{figure}[tbp]
\caption{Illustration of the experiment where the center of the
harmonic+optical trap, confining quasi-2D clouds of atoms, is suddenly
moved, which causes the oscillation of the superfluid. The grey wave
represents the magnetic+optical trapping potential and the ellipsoids are
the quasi-2D atomic clouds.}
\end{figure}
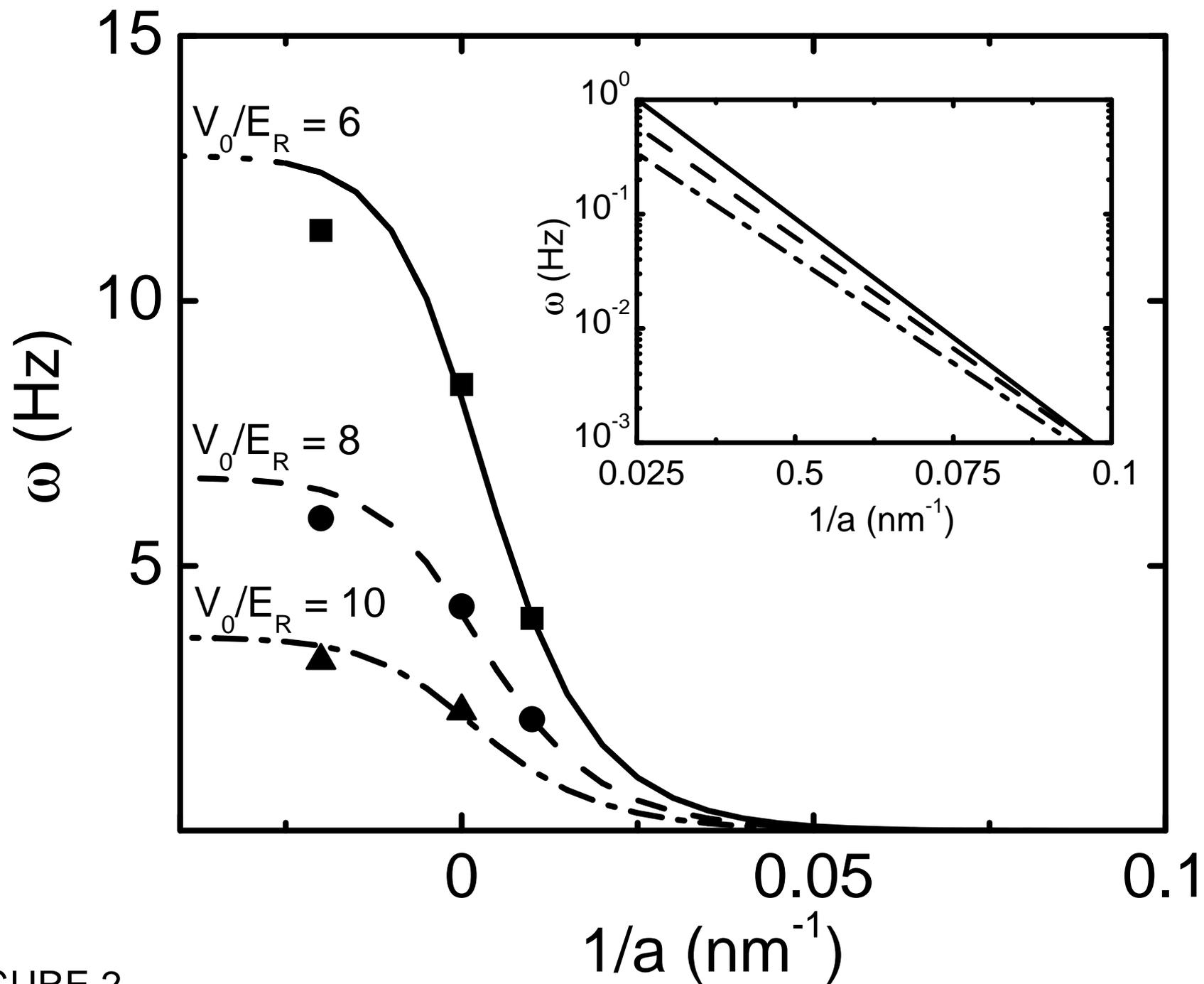
\begin{figure}[tbp]
\caption{The center of mass oscillation frequency of a zero temperature
fermionic superfluid in a one dimensional optical potential is shown as a
function of the inverse scattering length for different values of the
optical potential measured in units of the recoil energy $E_{R}$. The lines
are calculated with formula (\ref{wanal}) for $V_{0}/E_{R}=6$ (solid line), $%
V_{0}/E_{R}=8$ (dashed line), \ $V_{0}/E_{R}=10$ (dashed line). The left of
the curve $V_{0}/E_{R}=6$ is shown dotted because there the condition $%
t_{1}\ll \left\vert \Delta \right\vert $ is not fulfilled anymore. The
symbols show the frequency of a sinusoidal fit to the full numerical
solution of the equations (\ref{eqpsitot}) for $V_{0}/E_{R}=6$ (squares), $%
V_{0}/E_{R}=8$ (circles), \ $V_{0}/E_{R}=10$ (triangles).}
\end{figure}

\end{document}